\def\etal{{\it et al.\thinspace}}
\def\eg{{\it e.g.\ }}

\def\ie{{\it i.e.\ }}

\def\gsim{~\rlap{$>$}{\lower 1.0ex\hbox{$\sim$}}}

\documentclass[12pt,preprint]{aastex}

\begin{document}

\title{Stability Limits in Extra-solar Planetary Systems}

\author{Rory Barnes\altaffilmark{1,2} and Richard Greenberg\altaffilmark{1}}

\altaffiltext{1}{Lunar and Planetary Laboratory, University of Arizona,
Tucson, AZ 85721}
\altaffiltext{2}{For correspondence: rory@lpl.arizona.edu}

\keywords{methods: analytical, methods: N-body simulations, stars:
individual (HD 12661, 47 UMa), stars: planetary systems}

\begin{abstract} 
Two types of stability boundaries exist for any planetary system
consisting of one star and two planets.  Lagrange stability requires
that the planets remain bound to the star, conserves the ordering of
the distance from the star, and limits the variations of orbital
elements like semi-major axis and eccentricity. Hill stability only
requires that the ordering of the planets remain constant; the outer
planet may escape to infinity. A simple formula defines a region in
orbital element space that is guaranteed to be Hill stable, although
Hill stable orbits may lie outside the region as well. No analytic
criteria describe Lagrange stability.  We compare the results of 1000
numerical simulations of planetary systems similar to 47 UMa and HD
12661 with these two types of boundaries. All cases are consistent
with the analytic criterion for Hill stability. Moreover, the
numerically determined Lagrange boundary lies close to the analytic
boundary for Hill stability. This result suggests an analytic
formulation that may describe the criterion for Lagrange stability.
\end{abstract}

\section{Introduction}
Since the discovery of the first extra-solar planetary system with
multiple companions, $\upsilon$ And (Butler \etal 1999), substantial
research has investigated the dynamics of multiplanet systems. Most of
this work has examined the nature of individual systems such as 47 UMa
(Fischer \etal 2002; Laughlin, Chambers \& Fischer 2002; Go\'zdzeiwski
2002) and HD 12661 (Fischer \etal 2003; Go\'zdziewski 2003; Lee
\& Peale 2003). One investigation, using numerical integration of orbits, showed that several of the known systems lie near an
obvious stability boundary (Barnes \& Quinn 2004).

The dynamical stability of gravitational systems of multiple ($>2$)
particle systems has been studied for centuries. The description of
the motions in this type of system have no analytic solution. Analytic
constraints on dynamical stability began to emerge in the 1970's and
1980's, when it was shown that the motions of a system of two planets
and a star would be bounded in some situations (Zare 1977; Szebehely
1980; Marchal \& Bozis 1982; Milani \& Nobili 1983; Valsecchi, Carusi
\& Roy 1984). These constraints can be interpreted in terms of the
limitations on angular momentum exchange between the planets (Milani
\& Nobili 1983). However this type of argument is only valid for two planets not involved in any low-order mean motion resonances. There is no known
analytic boundary for systems in a low-order mean motion resonance, or
a system with more than two planets.

Two predominant definitions of stability have emerged.  In Hill (or
hierarchical) stability, the ordering of the planets, in terms of
distance from the central star, is conserved.  However, the outermost
planet may escape to infinity, and the system would still be
considered stable. A more useful definition, called Lagrange
stability, is more stringent: the planets remain bound to the central
star, changes in the ordering of the planets are forbidden, and the
semi-major axis and eccentricity variations also remain bounded.

Currently, investigations of the Lagrange stability of a system are
generally made through numerical simulations (\eg Barnes \& Quinn
2004).  However, Marchal \& Bozis (1982) noted that: ``Some studies
(Szebehely and McKenzie 1977; Szebehely, 1978, 1980) seem to show a
correlation between the Hill stability and the other types of
stability related to escape and exchanges; it would be interesting to
investigate these questions.'' In 1982 limited computer power made
such an investigation daunting. Now with modern computing power and
motivated by exoplanet systems we can revisit their supposition.

Gladman (1993) extended the study of Hill stability by approximating
the boundary (see $\S$2) in orbital element space. He verified the
analytic expression through numerical tests, in certain limits. More
recently, Veras \& Armitage (2004) modified the Hill criterion for application to mutually inclined orbits.

Other stability studies consider boundaries between periodic,
quasi-periodic and formally chaotic orbits via the
Fast Lyapunov Indicator (FLI) (Froeschl\'e, Lega, \& Gonczi
1997). Such boundaries have been
explored in extra-solar planetary systems (Go\'zdziewski \etal 2001;
Kiseleva-Eggleton \etal 2002; Go\'zdziewski 2002, 2003), however it is
not clear how these boundaries (or any other test based on Lyapunov
exponents) relate to limits of Lagrange stability.

In this paper, we compare the analytic description of \textit{Hill}
stability to a numerical determination of \textit{Lagrange}
stability. In $\S$2 we review the Hill stability equations for systems
of two planets. In $\S$3 we numerically test the analytic solutions
and compare the predictions of Hill stability with Lagrange stability,
which is determined by $N$-body simulations. In $\S$4 we draw general
conclusions and suggest directions for future work.

\section{Hill Stability}
There is no analytic solution for the motion of three
gravitating bodies, but in
certain situations the range of motion can be shown to be bounded; certain regions of
phase space are forbidden for each particle (Marchal \& Bozis 1982;
Roy \etal 1984; Milani \& Nobili 1983; Valsecchi, Carusi \& Roy
1984). This boundary is a direct result of the conservation of angular
momentum. For the case of two planets around a much more massive star,
the eccentricity exchange (through exchange of orbital angular momentum) is
limited, and the planets will never experience a close enough encounter to expel the interior planet from the system (\ie Hill stability).

Marchal \& Bozis (1982) quantified the criterion for Hill stable configurations as 
\begin{equation}
\label{eq:exact}
-\frac{2M}{G^2M_*^3}c^2h > 1 + 3^{4/3}\frac{m_1m_2}{m_3^{2/3}(m_1+m_2)^{4/3}} - \frac{m_1m_2(11m_1 + 7m_2)}{3m_3(m_1+m_2)^2} + ...,
\end{equation} 
where $M$ is the total mass of the system, $m_1$ and $m_2$ are the planet masses (the subscript 1 refers to the inner planet), $m_3$ is the mass of the star,  $G$ is the gravitational
constant, $M_* = m_1m_2 + m_1m_3 + m_2m_3$, $c$ is the total angular
momentum of the system, and $h$ is the energy. If a given
three-body system satisfies the inequality in Eq.\ (\ref{eq:exact}),
then the system is said to be Hill stable, and close approaches are
forbidden \textit{for all time}. If the inequality fails to be
satisfied, then the Hill stability of the system is unknown;
\textit{the system may still be Hill stable.}  Note that the left-hand side of Eq.\ (\ref{eq:exact}) is a function of
the positions and velocities of the system, and the right-hand side is purely a function of the masses. Thus for given masses, Eq.\ (\ref{eq:exact}) defines a boundary in orbital element space.

Gladman (1993) showed that Eq.\ (\ref{eq:exact}) could be changed to
barycentric orbital elements and rewritten, to first order, as
\begin{equation}
\label{eq:elements}
\alpha^{-3}\Bigg(\mu_1 + \frac{\mu_2}{\delta^2}\Bigg)(\mu_1\gamma_1 + \mu_2\gamma_2\delta)^2 > 1 + 3^{4/3}\frac{\mu_1\mu_2}{\alpha^{4/3}},
\end{equation}
where $\mu_i = m_i/M$, $\alpha = \mu_1 + \mu_2$, $\gamma_i = \sqrt{1 -
e_i^2}$, $\delta = \sqrt{a_2/a_1}$, $e$ is the eccentricity, $a$ is
the semi-major axis, and $i = 1, 2$. For given masses and
eccentricities, there is a critical value of
the semi-major axis ratio (or equivalently a critical value of
$\delta$, which we call $\delta_{crit}$), for which the two sides of
Eq.\ (\ref{eq:elements}) are equal. If $a_2/a_1$ is large enough (\ie
$\delta > \delta_{crit}$), then the system is surely Hill stable,
otherwise, maybe not.

The boundary for Lagrange
stability should lie at $\delta > \delta_{crit}$ (larger orbital
separation) because it is a more stringent definition of stability.
As we show below, this expectation is borne out by our numerical
integrations. There would be no reason to expect, \textit{a priori}, that the
actual Lagrange boundary would be correlated with the Hill boundary
limit. There might not even be a clear boundary in orbital element space between Lagrange
stable and Lagrange unstable configurations.

\section{Stability of Exoplanet Systems}
In this section we numerically explore the stability of hypothetical systems with masses and orbital elements similar to the 47 UMa and HD
12661 systems. In Table 1 we
present the current best fits (masses and orbits) and errors, shown in
parenthesis, for each of these two systems. In this table $m$ is the
planetary mass, $\varpi$ is the longitude of periastron, and
$T_{peri}$ is the time of periastron passage. Eqs. (\ref{eq:exact} -- \ref{eq:elements}) shuld apply to these systems because each has only two planets not in low-order mean
motion resonance.  Moreover, we can exploit orbital integrations that had
already been performed for different purposes (Barnes \& Quinn 2003,
2004). The numerical simulations were performed with MERCURY6
(Chambers 1999) for HD 12661 or SWIFT (Levison \& Duncan 1994) for 47
UMa.

For each of the two systems, Barnes \& Quinn (2003, 2004) considered
1000 different initial conditions distributed over the range of
observational uncertainty (Fischer \etal 2002; Fischer \etal
2003). For most orbital elements they selected values at random from a
Gaussian distribution. However the inclinations were selected from a
uniform distribution between 0 and 5$^o$, and the longitude of
ascending node from a uniform distribution from 0 to $2\pi$. For the
initial conditions each orbital element was selected
independently. This distribution is not ideal for mapping stability;
\textit{a priori}, a uniform distribution, far from any mean motion
resonances, might have been more efficient, except that we already had
these results in hand.

\subsection{HD 12661}
\begin{figure}
\plotone{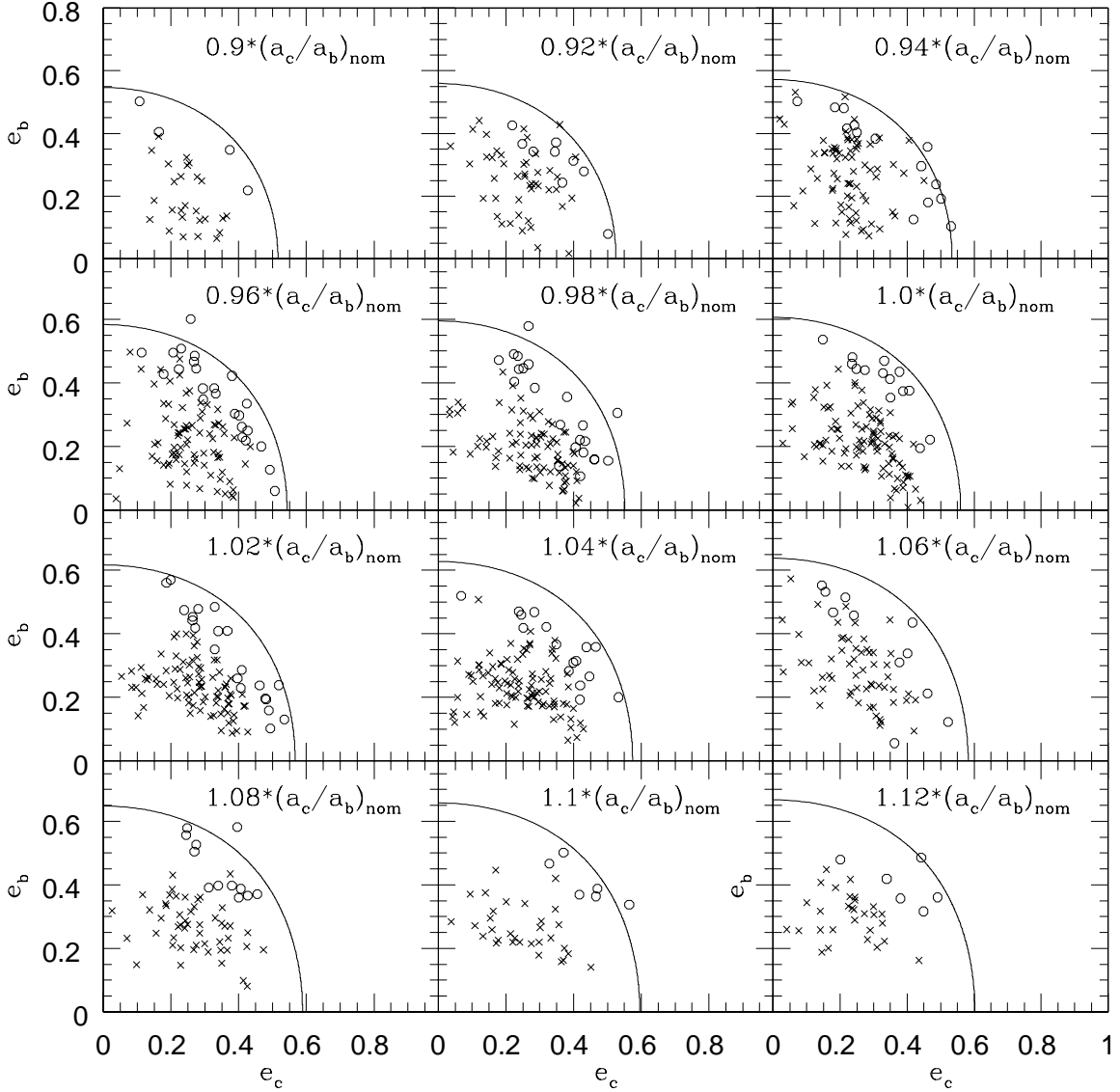}
\caption[]{\label{fig:slicehd12661} \small{Lagrange stability of different initial configurations of the HD 12661 planetary system based on $e_b$, $e_c$, and $a_c/a_b$. Each panel is a slice through this three dimensional space, showing all cases which begin with $a_c/a_b$ within 1\% of the value listed at the top of the panel. (Here the ratio $(a_c/a_b)_{nom}$ is the best fit ratio of the semi-major axes which is 3.08 for HD 12661. For
example, the top left panel contains all trials which began with a
ratio of $a_c/a_b$ in the range 0.89 to 0.91 times the best-fit
value.) x's represent Lagrange stable configurations, circles are Lagrange unstable, and the solid line repreents the solution to Eq.\ (\ref{eq:elements}).}}
\end{figure}

For HD 12661 the outcome after 4 million years of each numerical
experiment (Lagrange stability or instability) is shwon as a function
of $e_b$, $e_c$ and $a_c/a_b$ in Fig, \ref{fig:slicehd12661}. This
choice of timescale is somewhat arbitrary, but has been shown to
identify most unstable configurations (Ford, Havlickova
\& Rasio 2001; Barnes \& Quinn 2004). Also shown in Fig.\ \ref{fig:slicehd12661}, for comparison with the
numerical results, is the surface represented by Eq.\
(\ref{eq:elements}). According to Eq.\ (\ref{eq:elements}) all
configurations that lie to the lower left of the curves (smaller
eccentricities) \textit{must} be Hill stable. Note that this criterion is not exclusive: Hill
stable configurations are possible outside that region as
well. Therefore the actual boundary between Hill stability and
instability lies to the upper right of the curves. The Lagrange boundary must lie
below and to the actual Hill boundary of the curves because Lagrange stability is a
more stringent criterion.

In these results every case considered remained Hill stable over 4
million years (both x's and circles) consistent with the expectations
of Eq.\ (\ref{eq:elements}). Therefore, regardless of Lagrange
stability, these configurations were all Hill stable. In principle,
any case that is Hill stable and Lagrange unstable (the circles in
Fig.\ \ref{fig:slicehd12661}) could have gone Lagrange unstable either
by switching the planets' order or by ejecting the outer planet.
Every Lagrange unstable configuration of HD 12661 ejected the outer
planet (planet c). Most interestingly, the boundary of Lagrange
stability is close to, and tracks, the surface defined by Eq.\
(\ref{eq:elements}), which was derived in the context of Hill
stability. Marchal \& Bozis (1982) had suspected such a relationship.

\begin{figure}
\plotone{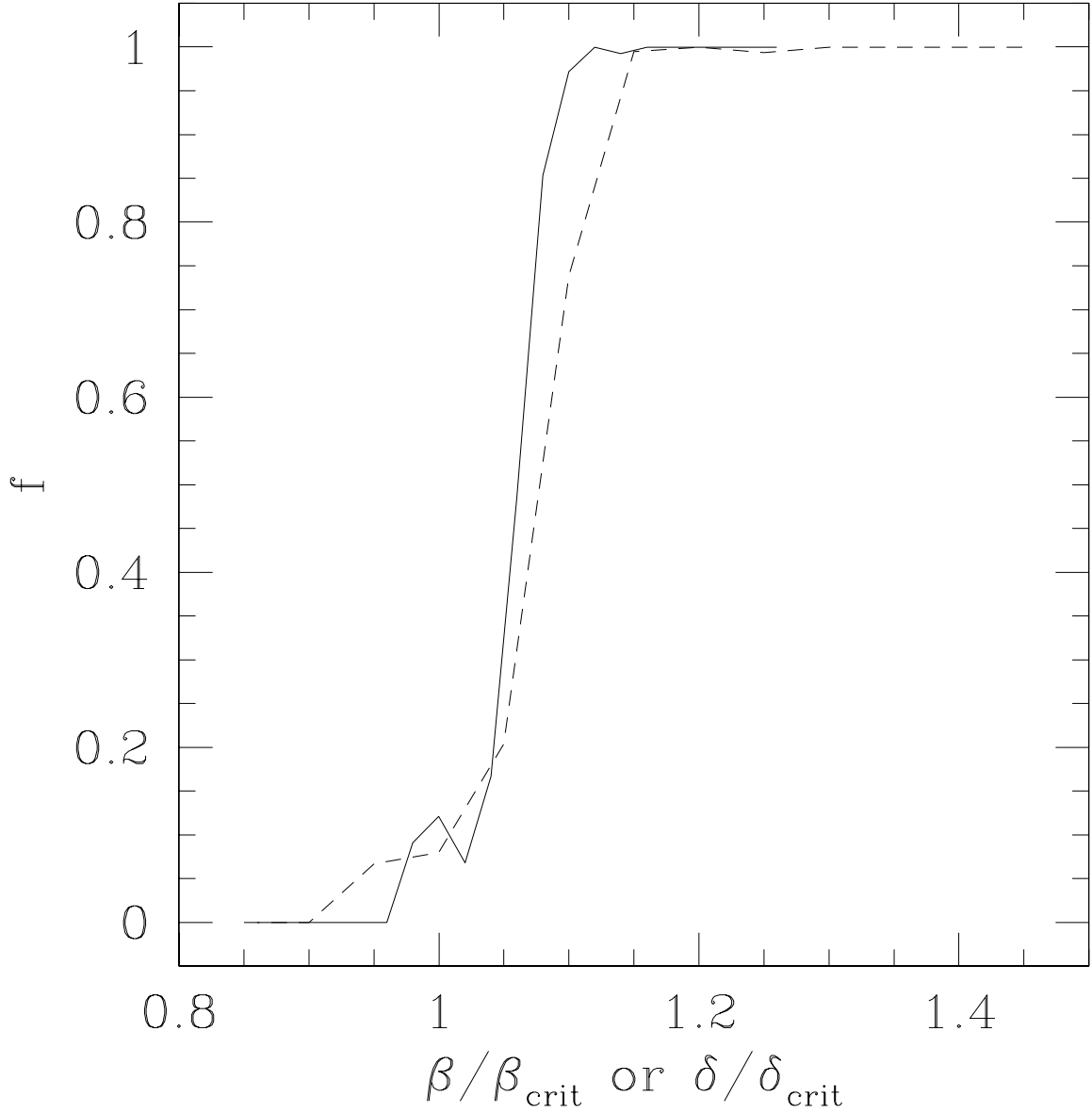}
\caption[]{\label{fig:xcrit} \small{Fraction of cases that are Lagrange stable as a function of proximity to the Hill stability boundary. In this plot we also compare of the exact solution to the Hill stability boundary, Eq.\ (\ref{eq:exact}) (solid line), to that of the approximate solution, Eq.\ (\ref{eq:elements}) (dashed line), for the simulations of HD 12661. The transition from instability to stability occurs at values slightly greater than 1. }}
\end{figure}

Next let us quantify how far the numerically determined Lagrange
boundary is from Eq.\ (\ref{eq:elements}). For each configuration we
determine the value of $\delta/\delta_{crit}$. We then plot as a function of  $\delta/\delta_{crit}$ the fraction, $f$,
in each bin that is Lagrange stable over 4 million years (Fig.\
\ref{fig:xcrit}). There is a sudden transition (independent of eccentricity) from
Lagrange unstable configurations to Lagrange stable near $\delta/\delta_{crit} = 1.05$. The Lagrange stability boundary lies close to the surface defined by Eq.\ (\ref{eq:elements}).

Eq. (\ref{eq:exact}) can also be compared with the results of our
numerical simulations. The left hand side of Eq.\ (\ref{eq:exact}) is a
function of orbital elements that we call $\beta$. Fig.\
\ref{fig:xcrit} also includes a plot of $f$ as a function of
$\beta/\beta_{crit}$ ($\beta_{crit}$ being the right hand side of Eq.\ (\ref{eq:exact})), showing a clear transition within about 5\% of
the boundary defined by Eq.\ (\ref{eq:exact}). Even though the initial
eccentricities may be large (some are over 0.5) the approximate
solution, Eq.\ (\ref{eq:elements}), appears to be in good agreement
with Eq.\ (\ref{eq:exact}). Both the $\beta$ and $\delta$ curves in
Fig.\ \ref{fig:xcrit} show there is a relatively narrow transition
from Lagrange stability to Lagrange instability.

For the best-fit values to the observed HD 12661 system, we find that
$\delta = 1.756$ and $\delta_{crit} = 1.476$. The ratio is 1.19,
putting this system within, but not deep within, the stable zone.

\subsection{47 UMa}
\begin{figure}
\plotone{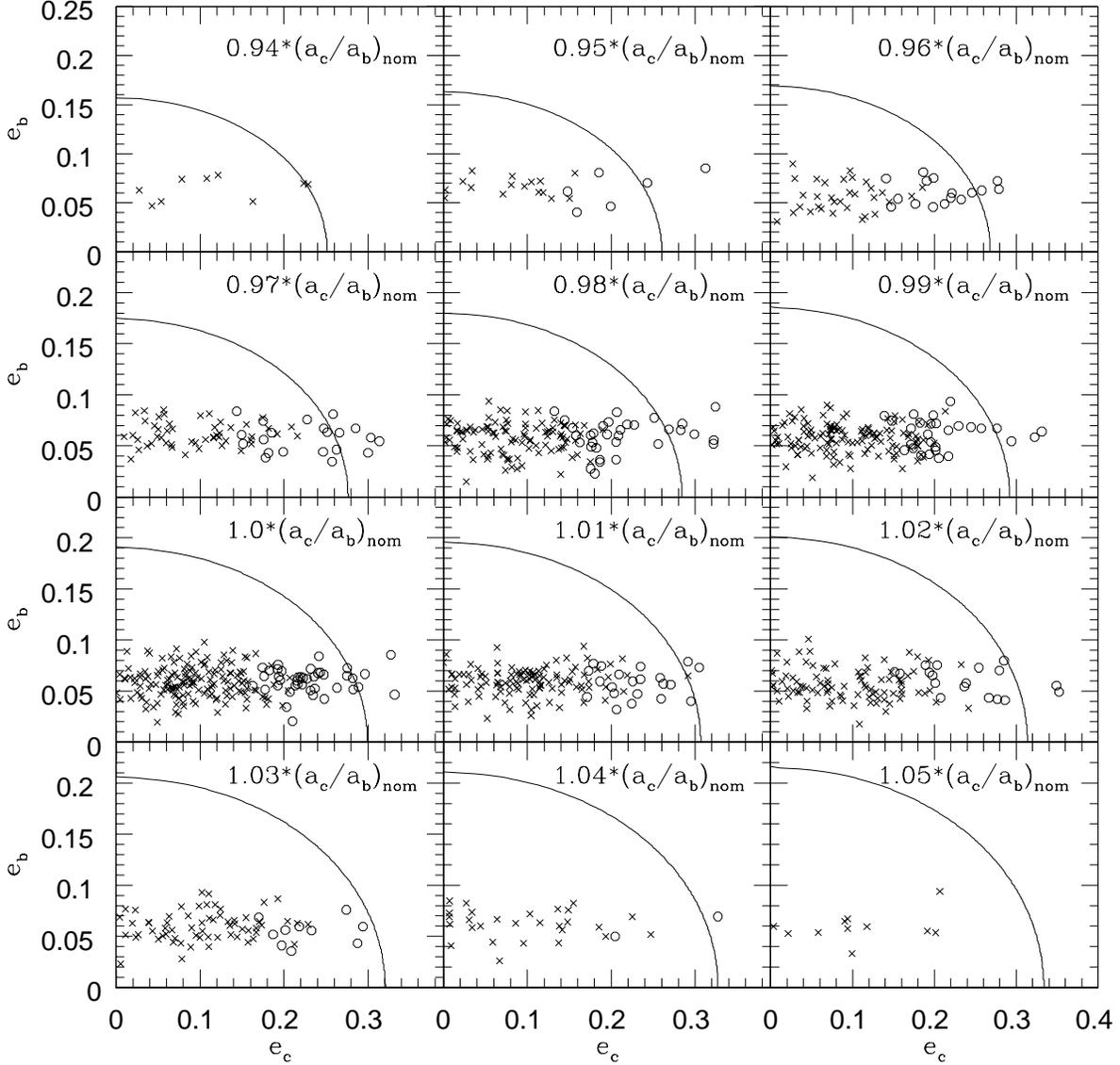}
\caption[]{\label{fig:slice47uma} \small{Lagrange stability of different initial configurations of the 47 UMa planetary system based on $e_b$, $e_c$, and $a_c/a_b$. The format is the same as Fig.\ \ref{fig:slicehd12661} except each panel is a slice through the parameter space showing cases which began within 0.5\% of the value of $a_c/a_b$ listed at the top of the panel. For 47 UMa the nominal ratio of the semi-major axes is 1.78. x's represent Lagrange stable configurations, circles are Lagrange unstable, and the solid line repreents the solution to Eq.\ (\ref{eq:elements}).}}
\end{figure}

Fig.\ \ref{fig:slice47uma} shows results for 47 UMa in a similar
format as Fig.\ \ref{fig:slicehd12661}. Results here are based on a
$10^6$ year timescale, which was shown to be a sufficient timescale to
determine stability (Barnes \& Quinn 2004). The eccentricity ranges
are different from one another (and from those in Fig.\
\ref{fig:slicehd12661}) because the uncertainties in the two
eccentricities are different (see Table 1). As in HD 12661, we see
that the Lagrange stability limit lies just inside the curve for Eq.\
(\ref{eq:elements}).

\begin{figure}
\plotone{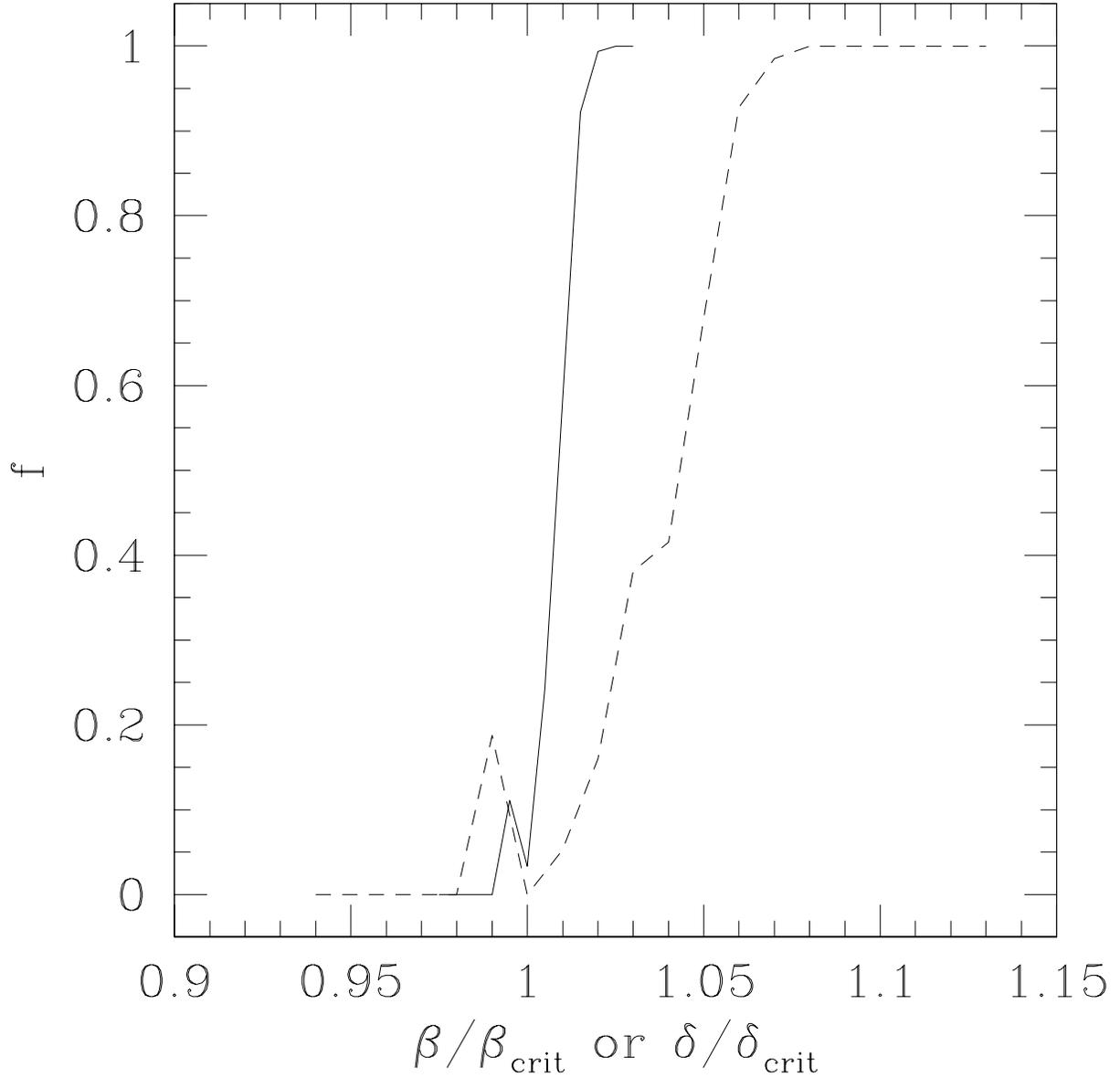}
\caption[]{\label{fig:47uma} \small{Lagrange stability of configurations of 47 UMa as a function of $\beta/\beta_{crit}$ (solid line) and $\delta/\delta_{crit}$ (dahsed line). As with the HD 
12661 system, Lagrange stability occurs at $\delta \approx 
1.1\delta_{crit}$.  }}
\end{figure}

Fig.\ \ref{fig:47uma} shows the fraction of Langrange stable
configurations (from Fig.\ \ref{fig:slice47uma}) as a function of
$\beta/\beta_{crit}$ and $\delta/\delta_{crit}$.  As with HD 12661,
the transition to Lagrange stability is at values of
$\beta/\beta_{crit}$ and $\delta/\delta_{crit}$ only slightly greater than
1.  Also, like HD 12661, every Lagrange unstable configuration ejected
the outer planet, confirming the Hill stability criterion. Once again
we see that the Lagrange stability boundary appears to track the
surface defined by Eq.\ (\ref{eq:elements}). Unlike HD 12661 the two
curves do not track each other exactly, but they are within a few percent,
consistent with the accuracy of the approximation of Eq.\
(\ref{eq:elements}).

For the best-fit values to the observed 47 UMa system, we find that $\delta = 1.336$ and $\delta_{crit} = 1.195$. Therefore the ratio of the two is 1.117, and the system is probably stable.

\section{Conclusions}
Although Eq.\ (\ref{eq:exact}), and its equivalent Eq.\
(\ref{eq:elements}), were derived in the context of Hill stability, we
have found that it appears to be a good predictor of Lagrange stability,
confirming the suspicions of Marchal \& Bozis (1982). At this point we tentatively conclude that if
$\delta \gsim 1.1\delta_{crit} \equiv \delta_{LS}$, then a two planet
system is Lagrange stable. In terms of $\beta/\beta_{crit}$, Langrange
stability appears to be guaranteed at slightly smaller
values. Additional work that numerically integrates various
hypothetical systems is needed to test the validity of these empirical
results. At this point, however, we tentatively find that if the ratio
of the semi-major axes were 1\% and 4\% closer for 47 UMa and HD
12661, respectively, then the Lagrange stability of the systems could
not be guaranteed. For these two systems, then, we have quantified how
far each is from the Lagrange stability boundary.

Eqs.\ (\ref{eq:exact}) and (\ref{eq:elements}) were derived in the
context of Hill stability, but they provide only weak
constraints. They do not actually define the boundary between Hill
stability and instability. Ironically, it now it appears that these
equations actually approximate the boundary of Lagrange stability.  We
would encourage a search for the explanation for this somewhat
surprising correlation between Eq.\ (\ref{eq:elements}) and the actual
Lagrange limit. If a physical explanation could be indentified, then
it may allow a quantification of Lagrange stability for an arbitrary
number of planets. Additionally it would be interesting to see if the
inclined Hill equation (Veras \& Armitage 2004) also tracks Lagrange
stability. Future numerical work may also determine how close Eq.\
(\ref{eq:elements}) is to the actual Hill limit.

The nature of Lagrange stability is a pressing issue given the
proximity of several systems to the boundary between Lagrange
stability and instability (Barnes \& Quinn 2001, 2004; Go\'zdziewski
2002, 2003). The proximities of these systems to Lagrange instability
have led to the Packed Planetary Systems hypothesis (Barnes \& Quinn
2004; Barnes \& Raymond 2004; Raymond \& Barnes 2005, Raymond, Barnes
\& Kaib 2006; see also Laskar 2000), which suggests that all planetary
pairs formed close to the Lagrange stability limit. The verification
or rejection of this hypothesis hinges on both theoretical advances
(such as this quantitative description of the Lagrange boundary) and
observational improvements (such as breaking the so-called
mass-inclination degeneracy and reductions in orbital element
errors). The results presented here might represent the first step
toward a theoretical understanding of the packing of planetary
systems.

This work was funded by NASA's Planetary Geology and Geophysics
program grant number NNG05GH65G. We would also like to thank an
anonymous referee for suggestions which greatly clarified this
manuscript.

\clearpage

\references
Barnes, R. \& Quinn, T. 2001, ApJ, 550, 884\\
--------. 2003 DDA ,\#34, \#8.01\\
--------. 2004 ApJ, 611, 494\\
Barnes, R. \& Raymond, S.N. 2004, ApJ, 617, 569\\
Butler, R.P. \etal 1999, ApJ, 526, 916\\
Chambers, J.E. 1999, MNRAS, 304, 793\\
Fischer, D. \etal 2003, ApJ, 586, 1394\\
Fischer, D. \etal 2002, ApJ, 564, 1028\\
Ford, E.B., Havlickova, M. \& Rasio, F. 2001 Icarus, 150, 303\\
Froeschl\'e, C., Lega, E. \& Gonzci, R. 1997, CeMDA, 67, 41\\
Gladman, B. 1993, Icarus, 106, 247\\
Go\'zdziewski, K. \etal 2001, A\&A, 378, 569\\
Go\'zdziewski, K. 2002, A\&A, 393, 997\\
--------. 2003, A\&A, 398, 1151\\
Kiseleva-Eggleton, L. \etal 2002, ApJL, 578, L145\\ 
Laskar, J. 2000, PhRvL, 84, 3240\\
Laughlin, G., Chambers, J. \& Fischer, D. 2002, ApJ, 579, 455\\
Lee, M.H. \& Peale, S.J. 2003, ApJ, 592, 1201\\
Levison, H.F., \& Duncan, M.J. 1994, Icarus, 108, 18\\
Marchal, C. \& Bozis, G. 1982, CeMech, 26, 311\\
Milani, A. \& Nobili, A.M. 1983, CeMech, 31, 213\\
Raymond, S.N. \& Barnes, R. 2005, ApJ, 619, 549\\
Raymond, S.N. \& Barnes, R. \& Kaib, N. 2006, ApJ, 644, 1223\\
Szebehely, V. 1978, CeMech, 18, 383\\
--------. 1980, CeMech, 22, 7\\
Szebehely, V. \& Mackenzie, R. 1977, AJ, 82, 79\\
Valsecchi, G.B., Carusi, B.A. \& Roy, A. 1984, CeMech, 32, 217\\
Veras, D. \& Armitage, P. 2004, Icarus, 172, 349\\
Zare, K. 1977, CeMech, 16, 35\\

\clearpage

\clearpage

\clearpage

\clearpage

\clearpage

\begin{center}Table 1: Orbital elements and errors\end{center}
{\small
\begin{tabular}{cccccccc}
\hline
System & $m_3$ (M$_\odot$) & Planet & $m$ ($M_{Jup}$) & $P$ (d) & $e$ & $\varpi$ ($^o$) & $T_{peri}$ (JD-2450000)\\
\hline
\hline
HD 12661 & 1.07 & b & 2.3 & 263.3 (20) & 0.35 (0.1) & 292.6 (20) & 9943.7 (10)\\
 & & c & 1.56 & 1444.5 (75) & 0.2 (0.1) & 147 (20) & 9673.9 (40)\\
47 UMa & 1.03 & b & 2.54 & 1089 (3) & 0.06 (0.014) & 172 (15) & 3622 (34)\\
 & & c & 0.76 & 2594 (90) & 0.1 (0.1) & 127.0 (56.0) & 1363.5 (493)\\ 
\end{tabular}
}

\end{document}